\documentclass{elsart}
\usepackage{graphicx,natbib,amssymb}
\journal{Physics Letters}

\newcommand{\beq}{\begin{equation}}
\newcommand{\eeq}{\end{equation}}
\newcommand{\lsi}{\,\raisebox{-0.13cm}{$\stackrel{\textstyle<}
{\textstyle\sim}$}\,}
\newcommand{\gsi}{\,\raisebox{-0.13cm}{$\stackrel{\textstyle>}
{\textstyle\sim}$}\,}
\newcommand{\be}{\begin{equation}}
\newcommand{\ee}{\end{equation}}

\begin{document}
\begin{frontmatter}

\title{\bf Non-binding of Flavor-Singlet\\ Hadrons to Nuclei}
\author{Glennys R. Farrar and Gabrijela Zaharijas}
\address{\it Center for Cosmology and Particle Physics, Department of Physics,\\
New York University, NY, NY 10003,USA}


\begin{abstract}
Strongly attractive color forces in the flavor singlet channel
may lead to a stable H dibaryon. Here we show that an H or other
compact, flavor singlet hadron is unlikely to bind to nuclei, so
that bounds on exotic isotopes do not exclude their stability.
Remarkably, a stable H appears to evade other experimental
constraints as well, when account is taken of its expected
compact spatial wavefunction.
\end{abstract}
\end{frontmatter}


\section{Introduction}
The spectrum of QCD may include a state of six quarks which is
simultaneously a singlet in color, flavor and spin, namely the H
dibaryon, with a quark content $uuddss$. It is a scalar with
charge 0 and strangeness -2, and is an isospin singlet and a
flavor singlet: $I(J^P)=0(0^+)$. In 1977, Jaffe predicted using
the bag model that the H would have a mass below $2 M_\Lambda$
\cite{jaffe} and thus be strong-interaction stable. Since then,
there have been many theoretical efforts to determine the mass
and production cross section of the H and, on the experimental
side, many inconclusive or unsuccessful attempts to produce and
detect it; see for example \cite{Hdibaryon}. An underlying
assumption has generally been that the H is not deeply bound.

In our work we examine the possibility that the H is lighter
than two nucleons, $m_H<2m_N$. The motivation comes from
phenomenological and theoretical analyses of QCD, as is detailed
in \cite{f:StableH}.  Briefly, the phenomenological argument
springs from the proposal that the puzzling properties of the
$\Lambda(1405)$ and $\Lambda(1520)$ are explained by their being
hybrid baryons consisting of a gluon bound to $uds$ quarks in a
flavor singlet-color octet state, to make an overall color
singlet\cite{kittel:Lam1405}. If the $\Lambda(1405)$ and
$\Lambda(1520)$ are hybrids and the glueball mass is $\sim 1.5$
GeV, a naive constituent quark model estimate leads to an H mass
in the range $\sim 1.3-1.5$ GeV\cite{f:StableH}. The other
approach is direct calculation using instanton liquid or
color-flavor locking arguments, which are known to imply strong
attraction in the diquark channel. Indeed, ref.
\cite{kochelev:H} states that an instanton liquid calculation
leads to $m_H = 1780$ MeV.

A tightly bound state generally is small in size.  In the
instanton liquid model this explains why $r_\pi < \frac{1}{2}
r_N$. Both instanton liquid and lattice calculations indicate
that the glueball is even more compact, so the H
$\leftrightarrow$ glueball analogy suggests $r_H \approx r_G
\lsi 1/4~r_N$.

In this and companion papers we explore the phenomenological
constraints on a stable H with mass in the range $1.3\lsi m_{H}
\lsi 2m_{p}$ and with radius $r_{H}\approx (1/6 - 1/4 )r_N$.
Elsewhere we show that such an H can be consistent with the
stability of nuclei and with $\Lambda$ decays in doubly-strange
hypernuclei \cite{fz:NucStab}. Here we investigate the binding
of the H, or more generally of any flavor singlet, to nuclei. We
determine the strength of coupling between the H and the
$\sigma$ meson or glueball which would be required for the H to
bind, and conclude that the H would not bind to nuclei if it is
as compact as hypothesized.  Thus the strong constraints on the
abundance of exotic isotopes do not exclude the existence of a
stable H.  If the H is stable and produced at the appropriate
level in the early Universe, it would be a good dark matter
candidate\cite{f:StableH}.  A mechanism which provides the
correct dark matter abundance will be described in
\cite{fz:HDM}.

In section \ref{expts} we summarize the relevant experimental
constraints on exotic nuclei. In section \ref{nucth}, we
summarize the theory of nuclear binding, to set the framework
for and to make clear the limitations of our computation. In
section \ref{calc} we analyze the binding of a flavor singlet
scalar to nuclei, and calculate the minimum values of coupling
constants needed for binding.  Corresponding limits on nucleon-H
scattering are given in section \ref{scat}.  Other
flavor-singlets are also considered, in section \ref{R0} and
elsewhere.  We summarize the results and give conclusions in
section \ref{conclusions}.

\section{Experimental constraints on the H binding}
\label{expts}

If the H binds to nuclei, experiments searching for anomalous
mass isotopes could be sensitive to its existence. Accelerator
mass spectroscopy (AMS) experiments generally have high
sensitivity to anomalous isotopes, limiting the fraction of
anomalous isotopes to $10^{-18}$ depending on the element. We
discuss binding of the H to heavy and to light isotopes
separately.

The H will bind more readily to heavy nuclei than to light ones
because their potential well is wider. However, searches for
exotic particles bound to heavy nuclei are limited to the search
for charged particles in Fe \cite{Feneg} and to the experiment
by Javorsek et al. \cite{javorsek} on Fe and Au. The experiment
by Javorsek searched for anomalous Au and Fe nuclei with $M_X$
in the range 200 to 350 atomic mass units u. Since the mass of
Au is 197 u, this experiment is sensitive to the detection of an
exotic particle with mass $M_X \ge 3$ u and is not sensitive to
the H with a mass $M_H \le 2$ u.

A summary of limits from various experiments on the
concentrations of exotic isotopes of light nuclei is given in
\cite{hemmick}. Only the measurements on hydrogen
\cite{hydrogen} and helium \cite{helium} nuclei are of interest
here because they are sensitive to the presence of a light
exotic particle with a mass of $M_X \sim~ 1 $ GeV. It is very
improbable that the H binds to hydrogen, since the $\Lambda$
does not bind to hydrogen in spite of having attractive
contributions to the potential not shared by the H, e.g., from
the $\eta$ and $\eta'$. Thus we consider only the limit on
helium. The limit on the concentration ratio of exotic to
non-exotic isotopes for helium comes from the measurements of
Klein, Middleton and Stevens who quote an upper limit of $\frac
{He_X}{He}<2\times 10^{-14}$ and $\frac{He_X}{He}<2\times
10^{-12}$ for primordial He \cite{plaga}.

\section{Nuclear binding-general}
\label{nucth}

QCD theory has not yet progressed enough to predict the two
nucleon interaction ab initio. Models for nuclear binding are,
therefore, constructed semi-phenomenologically and relay closely
on experimental input.

The long range part of the nucleon-nucleon interaction (for
distances $r\geq 1.5$ fm) is well explained by the exchange of
pions, and it is given by the one pion exchange potential
(OPEP). The complete interaction potential $v_{ij}$ is given by
$v^{\pi}_{ij}+v^{R}_{ij}$, where $v^{R}_{ij}$ contains all the
other (heavy meson, multiple meson and quark exchange) parts. In
the one boson exchange (OBE) models the potential $v^{R}_{ij}$
arises from the following contributions:
\begin{itemize}

\item In the intermediate region (at distances around $r\sim 1$
fm) the repulsive vector meson ($\rho ,\omega$) exchanges are
important.  A scalar meson denoted $\sigma$ was introduced to
provide an attractive potential needed to cancel the repulsion
coming from the dominant vector $\omega$ meson exchange in this
region. Moreover, a spin-orbit part to the potential from both
$\sigma$ and $\omega$ exchange is necessary to account for the
splitting of the $P^3$ phase shifts in NN scattering.

\item At shorter scales ($r\lsi 1$ fm), the potential is
dominated by the repulsive vector meson ($\rho ,\omega$)
exchanges.

\item For $r\lsi 0.5$ fm a phenomenological hard core repulsion
is introduced.
\end{itemize}
However, many of these OBE models required unrealistic values
for the meson-nucleon coupling constants and meson masses. With
this limitation the OBE theory predicts the properties of the
deuteron and of two-nucleon scattering, although, it cannot
reproduce the data with high accuracy.

A much better fit to the data is obtained by using
phenomenological potentials. In the early 1990's the Nijmegen
group \cite{nij90} extracted data on elastic NN scattering and
showed that all NN scattering phase shifts and mixing parameters
could be determined quite accurately. NN interaction models
which fit the Nijmegen database with a $\chi ^2/N_{data}\sim 1$
are called 'modern'. They include Nijmegen models \cite{nijmod},
the Argonne $v_{18}$ \cite{argonmod} and CD-Bonn
\cite{cdbonnmod} potentials. These potentials have several tens
of adjustable parameters, and give precision fits to a wide
range of nucleon scattering data.

The construction of 'modern' potentials can be illustrated with
the Nijmegen potential. That is an OBE model based on Regge pole
theory, with additional contributions to the potential from the
exchange of a Pomeron and f, f' and $A_2$ trajectories. These
new contributions give an appreciable repulsion in the central
region, playing a role analogous to the soft or hard core
repulsion needed in semi-phenomenological and OBE models.

Much less data exists on hyperon-nucleon interactions than on NN
interactions, and therefore those models are less constrained.
For example the extension of the Nijmegen potential to the
hyper-nuclear (YN) sector\cite{nijYN} leads to under-binding for
heavier systems. The extension to the $\Lambda \Lambda$ and $\Xi
N$ channels cannot be done without the introduction of extra
free parameters, and there are no scattering data at present for
their determination.

The brief review above shows that the description of baryon
binding is a difficult and subtle problem in QCD. Detailed
experimental data were needed in order to construct models which
can describe observed binding.  In the absence of such input
data for the H analysis, we must use a simple model based on
scalar meson exchange described by the Yukawa potential,
neglecting spin effects in the nucleon vertex in the first
approximation.  We know from the inadequacy of this approach in
the NN system that it can only be used as a crude guide. However
since the strength of couplings which would be needed for the H
to bind to light nuclei are very large, compared to their
expected values, we conclude that binding is unlikely. Thus
limits on exotic nuclei cannot be used to exclude the existence
of an H or other compact flavor singlet scalar or spin-1/2
hadron.

\section{Binding of a flavor singlet to nuclei}
\label{calc}

The H cannot bind through one pion exchange because of parity
and also flavor conservation. The absorption of a pion by the H
would lead to an isospin $I=1$ state with parity $(-1)^{J+1}$,
which could be $\Lambda \Sigma ^0$ or heavier $\Xi p$ composite
states. These states have mass $\gsi 0.7$ GeV higher than the
mass of the H, which introduces a strong suppression in $2^{nd}$
order perturbation theory. Moreover, the baryons in the
intermediate state must have relative angular momentum
$\rm{L}=1$, in order to have odd parity as required; this
introduces an additional suppression. Finally, production of
$\Lambda \Sigma ^0$ or $\Xi N$ states is further suppressed due
to the small size of the H, as explained in \cite{fz:NucStab}.
Due to all these effects, we conclude that the contribution of
one or two pion exchange to H binding is negligible.

The first order process can proceed only through the exchange of
a flavor singlet scalar meson and a glueball. The lightest
scalar meson is f(400-1100) (also called $\sigma$). The mass of
the glueball is considered to be around $\sim 1.5$ GeV.  In Born
approximation, the Yukawa  interaction leads to an attractive
Yukawa potential between nucleons \beq V(r)=-\frac {gg'}{4\pi}
\frac{1}{r}e^{-\mu r} \eeq where $\mu$ is the mass of the
exchanged singlet boson s ($\sigma$ or glueball) and $g g'$ is
the product of the s-H and s-nucleon coupling constants,
respectively. The potential of the interaction of H at a
position $\vec r$ with a nucleus, assuming a uniform
distribution of nucleon $\rho =\frac {A}{\rm V}$ inside a
nuclear radius R, is then \beq V=-\frac {gg'}{4\pi}\frac
{A}{{\rm V}} \int \frac{e^{ -\mu |\vec {r} -\vec {r'} |}}{|\vec
{r} -\vec {r'} |}d^3 \vec{r'} \eeq where A is the number of
nucleons, ${\rm V}$ is the volume of the nucleus and $\vec {r} $
is the position vector of the H. After integration over the
angles the potential is

\beq
V=-\frac {3}{2} \frac {gg'}{4\pi}
\frac{1}{(1.35~{\rm fm}~\mu)^3} f(r)
\eeq

where we used $R=1.35 A^{1/3}$ fm;
\begin{displaymath}
f(r) = \left \{ \begin{array}{ll}
 2 \mu \left[ 1 -(1+\mu R)~e^{-\mu R}~\frac{\sinh [\mu r]}{\mu r} \right]  & r\le R \\
2\mu \left[ \mu R\cosh [\mu R]-\sinh [\mu R] \right] \frac
{e^{-\mu r}}{\mu r} & r\ge R.
\end{array} \right.
\end{displaymath}
Throughout, we use $ \hbar = c = 1$ when convenient.

Figure 1 shows the potential the nucleus presents to the H for
A=50, taking the mass of the exchanged boson to be $\mu =$0.6
and 1.5 GeV. The depth of the potential is practically
independent of the number of nucleons and becomes shallower with
increasing scalar boson mass $\mu $.

\begin{figure}
\begin{center}
\includegraphics [width=8cm]{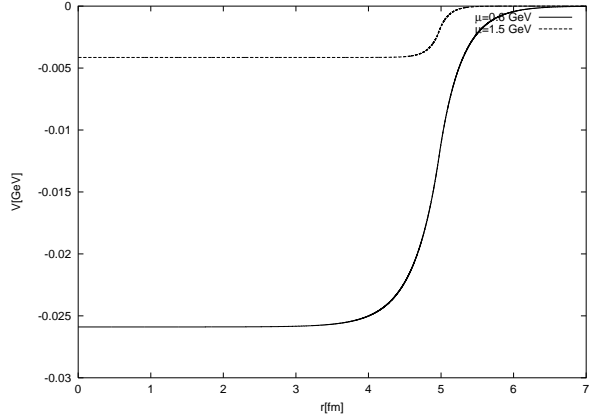}
\end{center}
\caption{Potential in GeV, for $\frac {gg'}{4\pi}$=1, A=50 and
$\mu=0.6~(\rm {dashed})$ or $\mu=1.5$ GeV (solid) as a function
of distance r.}
\end{figure}

Note that Born approximation is applicable at low energies and
for small coupling constants; it may not be valid for H binding.
Born approximation is valid when \beq \frac {m}{\mu } \frac
{gg'}{4\pi}<<1, \eeq where $m$ is the reduced mass and $\mu $
the mass of the exchanged particle. As we shall see, this
condition is actually not satisfied for values of $g g'$ which
assure binding for the H-mass range of interest.  This
underlines the fact that no good first principle approach to
nuclear binding is available at present.

We can now calculate the value of $c_*=\left( \frac
{gg'}{4\pi}\right) _*$ for which the potential is equal to the
minimum value needed for binding; in square well approximation
this is given by \beq V_{min}=\frac{\pi ^2}{8R^2m}. \eeq Figure
2 shows the dependence of $c_*$ on the mass of the exchanged
particle, $\mu$. The maximum value of $c_*$ for which the H does
not bind decreases with increasing H mass, and it gets higher
with increasing mass of the exchanged particle, $\mu$.

\begin{figure}
\begin{center}
\includegraphics*[width=8cm]{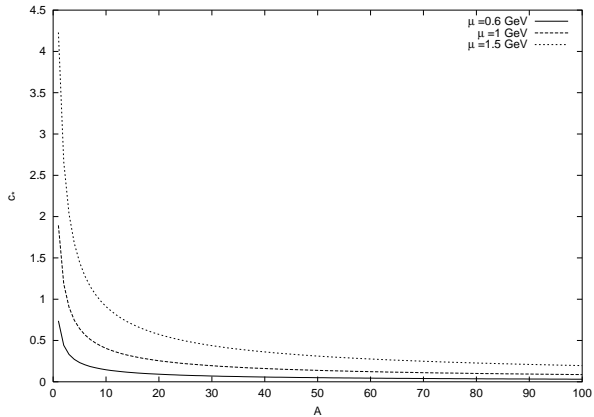}
\end{center}
\caption{ Critical value $c_*$ of the coupling constant product
versus nuclear size needed for the H to just bind, for
$\mu$[GeV] =0.7 (dotted), 1.3 (dashed) and 1.5 (solid).}
\end{figure}

The H does not bind to light nuclei with $A\le4$, as long as the
product of couplings $c_* \leq [0.27,0.73,1.65]$, for
$\mu=[0.6,1,1.5]$ GeV, where $c = g_{NN\sigma}~ g_{HH\sigma}/(4
\pi)$ or $g_{NNG}~ g_{HHG}/(4 \pi)$. The H will not bind to
heavier nuclei if $c_*\leq [0.019,0.054,0.12]$, for $\mu
=[0.6,1,1.5]$ GeV. In the next sections we will compare these
values to expectations and limits.

It should also be noted that binding requires the product of
coupling constants, $g g'$ to be positive and this may not be
the case. Even in the case of hyperons, experimental information
was necessary to decide whether the $\Xi$ has a relative
positive coupling \cite{dover}.

\section{Limits on {\bf {\it cm}} from Nucleon H scattering}
\label{scat}

The nucleon-H elastic scattering cross section is expected to be
very small, due to the compact size of the H and the suppression
of color fluctuations on scales $\lsi 1 ~{\rm GeV}^{-1}$ in the
nucleon.  Ref. \cite{f:StableH} estimates $\sigma_{HN} \lsi
10^{-3}$ mb.  This can be translated to an estimated upper limit
on the product $c ~m$ which determines the potential presented
to the H by a nucleus, as follows.  In the one boson exchange
model, the elastic H-N cross section due to the $\sigma$- or
glueball-mediated Yukawa interaction is given by \beq \frac
{d\sigma }{d\Omega }= (2mc)^2 \frac {1}{(2p^2(1-\cos \theta
)+\mu ^2)^2}. \eeq  In the low energy limit \beq
\label{eq:crossection} \sigma_{HN} =(2mc)^2\frac {4 \pi}{\mu
^4}. \eeq Writing $\sigma_{HN} = \sigma_{-3} 10^{-3}$ mb and
$\mu = \mu_{\rm GeV}$ 1 GeV, this gives \beq c~ m = 0.007
\sqrt{\sigma_{-3}} ~\mu_{\rm GeV}^2 ~{\rm GeV}. \eeq  Comparing
to the values of $c^*$ needed to bind, we see that for $m_H<2
m_p$ this is too small for the H to bind, even to heavy
nuclei\footnote{We have summarized the net effect of possibly
more than one exchange boson (e.g., $\sigma$ and glueball) by a
single effective boson represented by a $c_*^{\rm eff}$ and
$\mu_{\rm eff}$.}.

If dark matter consists of relic H's, we can demonstrate that
H's do not bind to nuclei without relying on the theoretical
estimate above for $\sigma_{HN}$. It was shown in \cite{wandelt}
that the XQC experiment excludes a dark matter-nucleon cross
section $\sigma_{XN}$ larger than about 0.03 mb for $m_X \sim
1.5$ GeV. Thus if dark matter consists of a stable H it would
require $\sigma_{XN} \le 0.03 $ mb, implying $c \le [0.01, 0.03,
0.06]$ for $\mu = [0.6,1.0,1.5]$ GeV and the H would not bind
even to heavy nuclei.

A generic new scalar flavor singlet hadron X which might appear
in an extention of the standard model, might not have a small
size and correspondingly small value of $\sigma_{XN}$, and it
might not be dark matter and subject to the XQC limit.  In that
case, it is more useful to turn the argument here around to give
the maximum $\sigma_{XN}^*$ above which the X would bind to
nuclei in the OBE approximation.  From eqn (3),(5) and $f(0) = 2
\mu$ we have

\beq c_* = \frac{\pi^2 (1.35 ~{\rm fm}) \mu^2}{24 A^{2/3} m}.
\eeq Then eqn (7) leads to \beq \sigma_{XN}^* \approx
155~A^{-4/3}~ {\rm mb}.\eeq  That is, for instance, if it is
required that the X not bind to He then it must have a cross
section on nucleons lower than 25 mb.

\section{Flavor singlet fermion}
\label{R0}

The analysis of the previous sections can be extended to the
case of a flavor singlet fermion such as the glueballino -- the
supersymmetric partner of the glueball which appears in theories
with a light gluino\cite{f:lightgluino}.  In this case the
possible exchanged bosons includes, in addition to the $\sigma$
and the glueball, the flavor-singlet component of the
pseudoscalar meson $\eta'$ ($m_\eta' = 958$ MeV).  However the
size of the $R_0$ should be comparable to the size of the
glueball, which was the basis for estimating the size of the H.
That is, we expect $r_{R_0} \approx r_G \approx r_H$ and thus
$\sigma_{R_0 N} \approx \sigma_{HN}$\cite{f:StableH}. Then
arguments of the previous section go through directly and show
the $R_0$ is unlikely to bind to light
nuclei\footnote{Nussinov\cite{nussinov:R0} considered that the
$R_0$ would bind to nuclei, by assuming that the depth of the
potential presented by the nucleus to the $R_0$ is at least 2-4
MeV for $16\le A\le 56$.  However the discussion of the previous
sections, with $\sigma_{R_0N} = 10^{-3}\sigma_{-3}$ mb, gives a
potential depth of 0.07 MeV $\sqrt{\sigma_{-3}}/(m_{R_0}/{\rm
GeV})$.}.

\section{Summary and Conclusions}
\label{conclusions}

As discussed in section \ref{expts}, experimental constraints on
the binding of a stable H or other flavor singlet scalar hadron
to nuclei are most restrictive for helium.  We reviewed the
theory of nuclear binding and emphasized that even for ordinary
nucleons and hyperons there is not a satisfactory
first-principles treatment of nuclear binding.  We showed that
exchange of any pseudoscalar meson, or of two pseudoscalar octet
mesons, or any member of the vector meson octet, makes a
negligible contribution to the binding of an H or other flavor
singlet scalar hadron to a nucleon.  The dominant attractive
force comes from exchange of a glueball or a $\sigma$ (also
known as the f(400-1100) meson), which we treated with a simple
one boson exchange model.  The couplings of $\sigma$ and
glueball to the $H$ are strongly constrained by limits on
$\sigma_{HN}$, to such low values that the H cannot be expected
to bind, even to heavy nuclei.

Thus we conclude that the strong experimental limits on the
existence of exotic isotopes of He and other nuclei do not
exclude a stable H. More generally, our result can be applied to
any new flavor singlet scalar particle X, another example being
the $S^0$ supersymmetric hybrid baryon ($uds\tilde{g}$)
discussed in \cite{f:lightgluino}. If $\sigma_{XN} \le 25~{\rm
mb} ~{\rm GeV}/m_{X}$, the X particle will not bind to light
nuclei and is ``safe". Conversely, if $\sigma_{XN} >> 25~{\rm
mb} ~{\rm GeV}/m_{X}$, the X particle could bind to light nuclei
and is therefore excluded unless it there is some mechanism
suppressing its abundance on Earth, or it could be shown to have
an intrinsically repulsive interaction with nucleons. This means
the self-interacting dark matter (SIDM) particle postulated by
Spergel and Steinhardt\cite{spergel:SIDM} to ameliorate some
difficulties with Cold Dark Matter, probably cannot be a hadron.
SIDM requires $\sigma_{XX} /M_X \approx 0.1 - 1 $ b/GeV; if X
were a hadron with such a large cross section, then on geometric
grounds one would expect $\sigma_{XN} \approx 1/4 \sigma_{XX}$
which would imply the $X$ binds to nuclei and would therefore be
excluded by experimental limits discussed above.

{\bf Acknowledgements}  The research of GRF was supported in
part by NSF-PHY-0101738.  GZ would like to thank Emiliano
Sefusatti for useful discussions.

\bibliographystyle{unsrt}
\bibliography{bindingH}

\end{document}